# A resistive electron irradiation microsensor made from conductive electrospun polycaprolactone fibers loaded with carbon nanotubes and fullerene C60

Fabricio N. Molinari[1,2], Maria A. Mancuso[2], Emanuel Bilbao[3], Theo Rodríguez Campos[3,4], Gustavo Giménez[3], Leandro N. Monsalve[3,4,5]

[1] INTI Textiles, Av. Gral. Paz 5445, San Martín, Buenos Aires, Argentina

[2] Institute of Atmospheric Pollution Research—National Research Council (IIA—CNR), Research Area of Rome 1, Strada Provinciale 35d, 9-00010 Montelibretti, Italy

[3] INTI Micro y Nanotecnologías, Av. Gral. Paz 5445, San Martín, Buenos Aires, Argentina

[4] Instituto de la Calidad Industrial (INCALIN – INTI – UNSAM), Av. Gral. Paz 5445, San Martín, Buenos Aires, Argentina

[5] Consejo Nacional de Investigaciones Científicas y Técnicas (CONICET), Godoy Cruz 2290, CABA, Argentina

**Abstract**

In this work an electron radiation detector microdevices were fabricated and characterized. The devices consisted of a conductive electrospun mat made of polycaprolactone loaded with multiwalled carbon nanotubes and fullerene C60 deposited onto gold interdigitated microelectrodes. They were capable of permanently increase their conductivity upon exposure to electron beam irradiation from 0.02 pC/µm$^2$ accelerated at 10 and 20 keV. This phenomenon could be explained due to the ability of C60 to trap and stabilize negative charges and thus contribute to the conductivity of the polymer composite. The devices achieved their maximum conductivity at an irradiation between 0.22 and 0.27 pC/µm$^2$ and this maximum was dependent of the electron acceleration. Montecarlo simulations were performed to explain dependence as function of electron penetration in the polymer composite. Moreover, the devices irradiated at 20keV maintained their final conductivity and the devices irradiated at 10keV increased their final conductivity after 6 days from irradiation. Fullerenes proved to act as highly efficient electron scavengers within the polymer composite and contribute to its conductivity, and the microdevice has potential application as beta radiation sensors.

1. Introducción

C60/polymer composites are well-known electron scavengers due to the electronic properties of C60[1]. C60 has a strong electron affinity and acts as a *n*-type semiconductor when used as channel in field effect transistors[2]. Moreover, charge injection in C60/polymer composites can cause ionization of C60, which enhances the electrical conductivity of the composite. This feature has been employed for the fabrication of C60-based memory devices[3–6].

Electron beams are generated through acceleration and collimation of electrons from an electron source using strong electric fields in vacuum. They have multiple applications: electron microscopy, welding, sterilization and radiotherapy among others. For the case of radiotherapy, calibration of the beam in terms of the dose administrated to a patient is critical in order to ensure the safety and efficacy of the treatment. Normally, different types of dosimeters are employed for the calibration of electron beams: polymer gels[7–10] or radiochromic films[11] for less accurate measurements and alanine films or pellets for more accurate measurements[12,13]. However, the response of these dosimeters has to be measured using spectrophotometry or electron paramagnetic resonance, which makes continuous monitoring of radiation during radiotherapy difficult. On the other hand, electronic microdosimeters whose primary signal is electrical are much more suitable for continuous monitoring and ensuring homogeneous irradiation. For instance, commercial MOSFET dosimeters were successfully employed for monitoring electron beam irradiation for intraoperative radiation therapy[14,15]. MOSFETs can serve as dosimeters since charges generated and trapped within the gate insulator by radiation shifts permanently their threshold volage ($V_T$). However, the application of MOSFETs for electron radiation therapy dosimetry is limited by their cost and reproducibility. Another way for real-time monitoring of electron beams is the use of devices made of polymers or polymer composites that change their electrical properties upon irradiation[16].

Based on our previous work involving the fabrication of memory devices based on electrospun polycaprolactone with carbon nanotubes and C60[6], we decided to test their ability to sense electron radiation and investigate their potential application as low-cost and reproducible dosimeters.

## 2. Materials and Methods

**Materials.** PCL (Mw 80000, Sigma-Aldrich), Fullerenes C60 (99.9%, Sigma-Aldrich), polyvinyl pyrrolidone PVP K30 (Anedra, Argentina), multiwalled carbon nanotubes (MWCNT) (Nanocyl 7000, Belgium), xylene, toluene, hexane, DMF, and acetone were reagent grade and used straight from the bottle.

**Preparation of Polymer Solutions.** Fullerene C60 (35 mg) was dissolved in xylene (11 ml) in ultrasound bath at 40 °C for 90 minutes. Then 2.7 g of PCL was added to 10ml of the C60-xylene solution and dissolved at 50 °C under magnetic stirring. The resulting solution was mixed with 10 g of a dispersion of MWCNTs in DMF (0.7%wt.) prepared according to a previously described procedure[17]. Once homogeneous additional gram of PCL was dissolved under magnetic stirring for 3 h and used immediately.

**Electrospinning conditions and Device fabrication.** A Y-flow electrospinner 2.2.D-500 (Y-flow SD, Spain) was used for electrospinning. Distance to the collector was optimized in order to obtain regular and dry fibers. Different tests determined 26 cm as the optimum distance between the needle and the collector. The flow rate of the solution was 1 ml.h$^{-1}$ for all samples. A rotary drum collector was employed to produce mats of aligned fibers. The setup included two high voltage sources: one at the needle between +6 and +12 kV and the other at the collector between -15 and -17 kV. A rotation speed of 500 rpm was selected for the collector in order to obtain aligned fibers.

PCL-MWCNT/C60 solutions were electrospun onto interdigitated sputtered gold electrodes of different width to length (W/L) ratio (L=10-50 µm, W=500-10.000) patterned on a Si/SiO$_2$ (300 nm) substrate for 15 minutes using a rotary collector to align the fibers perpendicular to electrode fingers (Figure 1). The dies with the deposited fibers were annealed at 60 °C for 20 minutes to improve contact between fibers and electrodes.

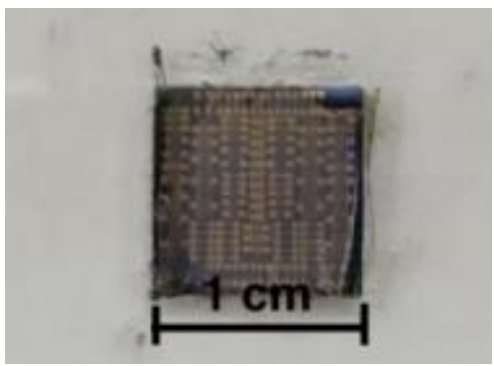

**Figure 1.** 1cm x 1cm Si/SiO$_2$ die with interdigitated microelectrodes and electrospun fibers.

**Characterization**

Scanning electron microscopy and Focused Ion Beam (FIB) experiments were performed in a Helios Nanolab 650 (FEI). Fiber diameter, alignment and surface coverage measurements were carried out by analysis of SEM images using Image J software[18]. Electrical characterization was performed using a Keithley 4200 SCS equipped with a manual probe station (Wentworth Lab AVT 702).

**Electron irradiation**

Electron irradiation was performed in a Helios Nanolab 650 (FEI). Doses (*D*) were calculated according to recommendations published by Egerton[19]. Devices of different areas (0.018 – 0.625 mm$^2$) were selected individually and irradiated with a fixed current of 0.8nA ($I_b$) for three minutes (*t*), and scan frequency of 300ns (*f*) and a Dwell time of 35ns (*Dw*) for three minutes in a given irradiation area (*A*) so that they all received a charge of 15.8416nC. The dose was given as a function of the area of each of the devices according to Equation 1:

$$D = \frac{I_b \cdot t \cdot Dw}{A \cdot f}$$

In these experiments it is impossible to define an exact dose, so the dose was time controlled with the previously cited parameters. A very soft imaging condition were used (1KV and 50pA) in order to minimize the irradiation in the visualization of each device, so that the penetration is minimal and proceeded as quickly as possible (Reference bar of Figure 1) to choose the irradiation area, since it is necessary to irradiate the sample in order to visualise visualize the devices, focus and select the area of irradiation.

**Results and Discussion**

**Fiber characterization**

SEM images of devices were taken to characterize fiber morphology (Figure 2). These images revealed a monolayer of fibers aligned perpendicular to electrode fingers. Regular fibers with smooth surface were observed. It was observed that the thermal treatment performed in order to optimize electrical properties of the device did not affect the fibrous structure of the electrospun layer.

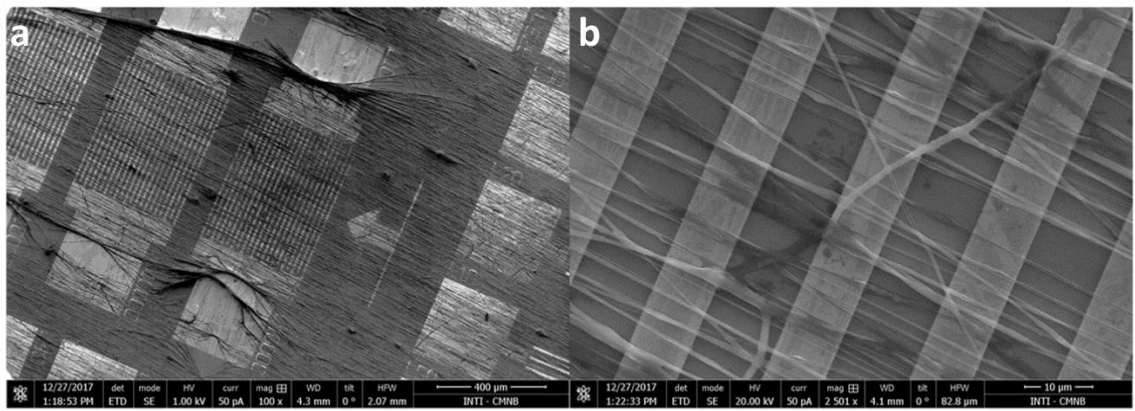

**Figure 2.** SEM images of PCL-MWCNT/C60 electrospun fibers deposited onto interdigitated electrodes at 100x (a) and 2500x (b).

Diameter and angle respect to the electrode fingers distribution were analyzed and the results are summarized in Table 1.

**Table 1.** Average diameter and angle of electrospun PCL-MWCNT/C60 fibers.

| | |
|---|---|
| **Fiber diameter (nm)** | 650 ± 18 |
| **Angle respect to the electrode fingers (°)** | 91.3 ± 0.4 |

**Surface coverage of the electrospun composite onto the electrodes**

Surface coverage was analyzed using color threshold tool of the image J software.[18] This tool allowed us to determine the area that fibers are occupying in the image compared to the total area of the image. The calculation of areas and porosity are shown in Table 2.

**Table 2.** Analysis of surface coverage.

| | |
|---|---|
| **Total area of image** | 67,374 μ$^2$ |
| **Fibers' area** | 13,226 μ$^2$ |
| **Surface coverage** $\left[\frac{Fibers\ Area}{Total\ Area} x 100\right]$ | 20% |

Regarding their nanostructure, the fibers contain a dispersion of MWCNTs decorated with C60. These C60/MWCNT complexes are responsible of their electrical switching properties and their nanostructure has been determined in a previous publication[6].

## Electrical properties

As stated in our previous work, these composite fibers reveal a resistive switching effect associated with an accumulation of negative charge in the C60/MWCNT complex. The device was stimulated by passing a current through it generating a decrease in the device resistance which is dependent on the bias voltage[6]. Surface resistivity of the fibers is 4.94 x 10$^8$ Ω/sq. When used as a memory device and a programming voltage was applied through the electrodes their resistance decreased down to 25% of its original value.

In this work, the possibility of stimulating the devices by irradiation with an electron beam is explored. For this purpose, an experiment was designed using a scanning electron microscope as a source of electrons and the devices as targets. During the experiment, devices fabricated using interdigitated electrodes with incremental active area were irradiated using accelerated electrons at 10kV and 20kV.

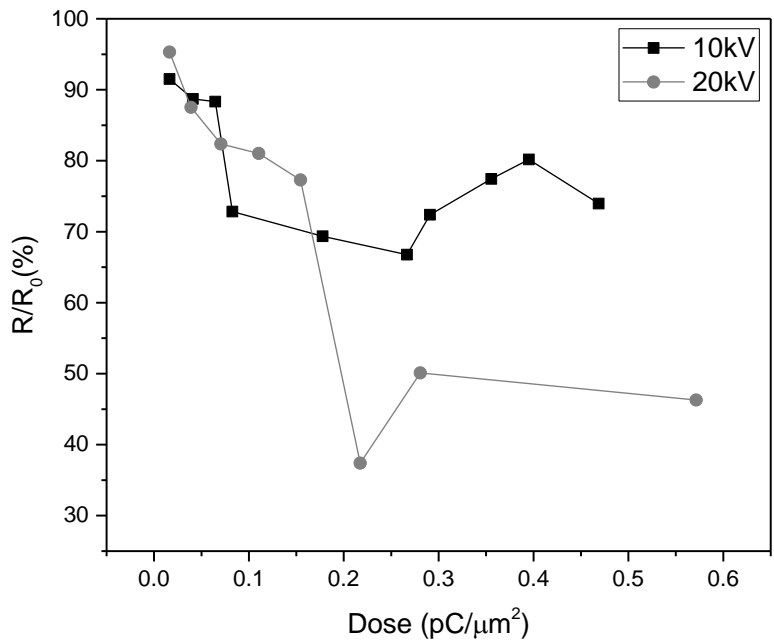

**Figure 3.** Relative resistance referred to the resistance of devices before irradiation as function of dose of electrons accelerated at 10kV (black squares) and 20kV (grey circles).

Figure 3 shows the variation of the resistance of the devices for different doses applied with accelerated electrons at 10kV and 20kV. It could be noticed that all the devices present an $R/R_0$ less than 1 regardless of the dose applied, which means that, in all cases, there was a decrease in electrical resistance after irradiation. Furthermore, it was observed that irradiation at 20kV generates a greater resistance variation than 10kV accelerated electrons. This phenomenon demonstrates that, since the number of electrons impacted per unit area is the same, the acceleration mainly affects the penetration of the electrons into the matrix.

Figure 4 shows the Monte Carlo simulations using the open-source software Casino V2.5[20] that were carried out in order to obtain the hypothetical penetration of electrons in the fibers. Fiber density was estimated from the chemical formula of the repeating unit of PCL ($C_6H_{10}O_2$). In the simulation, it was noticed that the electrons are able to penetrate the whole layer with 50% of their initial energy at 20kV of acceleration (a monolayer of fibers has an average diameter of 650nm), while they penetrate 260nm at 50% of the initial energy at 10kV. Electrons can trespass the whole layer only with an average 10% of their initial energy.

By penetrating further and with greater energy, electrons with sufficient energy can be trapped by the C60/MWCNT complexes that are found within the fibers not only at the surface but in the whole fibers' volume. Electrons reaching by C60 molecules allow the formation of new conduction channels and produce a further decrease in electrical resistance. Another highlighted point was that both curves had a similar shape, reaching a minimum resistance for a dose of 0.27pC/µm2 and then showing an increase in resistance for higher doses. This increase in resistance could be associated with over-stimulation of the device. This effect was also observed in devices powered with voltage cycles reported previously[6]. Probably, the super accumulation of charges blocks/breaks some conduction channels, thus generating an increase in resistance.

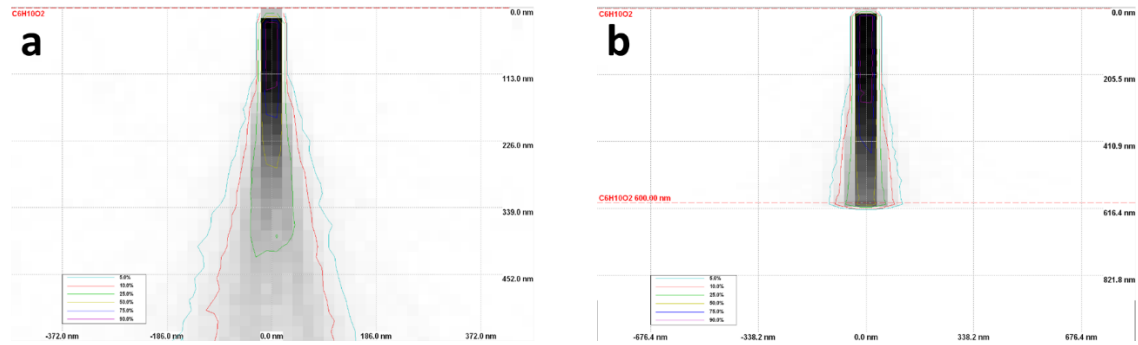

**Figure 4.** Simulated electron beam penetration in PCL fibers at 10kV (a) and 20kV (b).

In order to evaluate the stability over time of the switching induced by an electron beam, the resistance value on the devices was measured after 2 and 6 days from irradiation. Figure 5 shows the evolution of the relative resistance of the devices as a function of electron dose and time after irradiation.

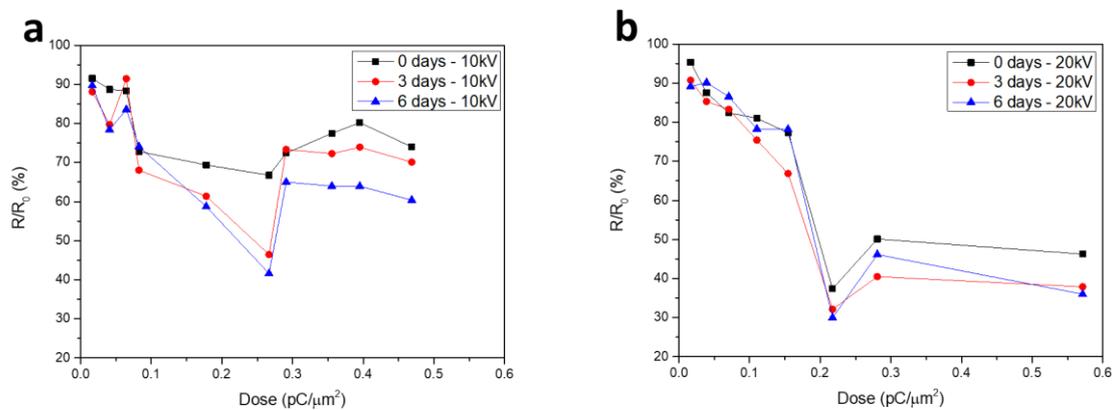

**Figure 5.** Relative resistance referred to the resistance of devices before irradiation as function of dose and time after irradiation of electrons accelerated at 10kV (a) and 20kV (b).

Devices irradiated at 10 kV lowered their resistance over time after irradiation event. This decrease in electrical resistance was more pronounced at a dose of 0.27 pC/μm$^2$ after 6 days from irradiation, which was 70% of its original value immediately after irradiation and reached 40% after 6 days. Interestingly, this effect was much less pronounced when the devices were irradiated at 20kV. This phenomenon could be related to electron beam penetration. Charge accumulation in the fibers takes place during irradiation and it's proportional to the dose. However, in order to impact in the electric resistance these charges have to be scavenged by C060 molecules. When irradiated at 10 kV, the electrons do not have sufficient energy to trespass the fiber width. Some of them are trapped by C60 leading to a decrease in electrical resistance and the remaining electrons produces ionization in the PCL molecules with no immediate impact in electrical properties. However, these ionized PCL molecules are able to migrate within the polymer matrix and transfer their charge to C60 molecules not previously affected by the electron beam leading to a secondary reduction of electrical resistance within days after electron irradiation. On the other hand, devices irradiated at 20kV did not show such large change in electrical resistance over time after irradiation as electron beam is able to penetrate the whole volume of the fibers and ionize C60 molecules more effectively. This hypothesis is supported by dielectric relaxation experiments carried out by Grimau et al. on PCL. Authors showed that, at 30°C, quasistatic charges have sufficient mobility within the polymer to migrate and rearrange[21].

Another feature of the devices is that their electrical resistance reaches a minimum at a dose of 0.27 pC/μm$^2$ and 0.22 pC/μm$^2$ at 10 and 20kV respectively. The minima for R/R$_0$ was 41% and 30% at 10 and 20kV respectively, which is consistent with our previous results regarding resistive switching behavior of the same PCL-C60/MWCNT electrospun nanocomposite. These minima can be regarded as saturation points at which every C60 molecule was fully ionized. The ratio between the electrons and C60 molecules can be estimated as follows: The quantity of irradiated electrons ($n_e$) in a given area $A$ is calculated by dividing the dose ($D_0$) by the charge of a mole of electrons ($Q$) (equation 2):

$$\frac{n_e}{A} = \frac{D_0}{Q} \qquad (2)$$

This gives between 2.28 and 2.80 x $10^{-18}$ moles of electrons/$\mu m^2$ for achieving minimum electrical resistance.

On the other hand, the molecular density of C60 ($n_{C60}/A$) in the electrospun composite can be calculated from the mass fraction of C60 ($C=0.8\%$), the density of the composite ($\delta=1.145$ x $10^{-12}$ g/$\mu m^3$, estimated from the density of PCL), the thickness of the fibers' layer ($E=0.65\mu m$, estimated from the average thickness of a single fiber), the surface coverage of the fibers' layer ($P=20\%$) and the molecular weight of C60 ($Mr_{C60}=7.21$ x $10^2$ g/mol) (equation 3):

$$\frac{n_{C60}}{A} = E * \delta * \frac{C \cdot P}{10.000 \cdot Mr_{C60}} \qquad (3)$$

Which gives 1.65 x $10^{-18}$ moles of C60/$\mu m^2$.

According to simulations conducted by Qui et al. on C60 molecular wires, the number of electrons per C60 needed for achieving a decrease of electrical resistance was 3.5 in 4-molecule-long wires and 2.7 for 10-molecule-long wires[22]. In our case, the ratio of electrons to C60 at the maxima decrease of electrical resistance was between 1.38 and 1.7. This fact demonstrates that there is a quantitative reaction between the electrons and C60 being the scavenging yield of C60 very high. These devices are very efficient reacting with incident electrons and have potential application for monitoring beta radiation during radiotherapy, similar to commercial MOSFET sensors [14,15].

In the case of these microdevices made of electrospun polymer composite fibers, they have the potential advantage of exposing many devices to a single beam and having a sensing layer of similar density and composition to biological tissue, thus allowing precise and in vivo monitoring of radiation dose and uniformity.

## 3. Conclusions

Fibers of PCL with C60-MWCNT electrospun nanocomposite were used for the fabrication of beta radiation microsensors based on the change of its electrical properties upon electron scavenging of C60. The devices were irradiated using an electron scanning microscope. They were sensitive to electron irradiation, showing a minimum of electrical resistance proportional to electron acceleration at 10kV and 20kV. For electrons irradiated at 10kV the electrical resistance decreased over time after irradiation and this effect was due to incomplete penetration of the electron beam across the fibers and charge reconfiguration due to the mobility of polymer chains. Both devices have shown a minimum $R/R_0$ for doses between 0.22 and 0.27pC/µm$^2$. The ratio of incident electrons to C60 molecules at minima of electrical resistance was calculated between 1.38 and 1.7. This demonstrates that PCL/C60-MWCNT electrospun fiber are very efficient as electron scavengers and are potentially useful as an active material for the fabrication of microdevices sensitive to beta radiation for application to monitoring radiation dose and homogeneity on radiotherapy.